# Site-Controlled Telecom Single-Photon Emitters in Atomically-thin MoTe$_2$

*Huan Zhao[*], Michael T. Pettes, Yu Zheng, and Han Htoon[*]*

Center for Integrated Nanotechnologies, Materials Physics and Applications Division, Los Alamos National Laboratory, Los Alamos, New Mexico 87545, USA

[*]Corresponding authors, Email: huanzha@lanl.gov, htoon@lanl.gov

**Quantum emitters (QEs) in two-dimensional transition metal dichalcogenides (2D TMDCs) have advanced to the forefront of quantum communication and transduction research[1] due to their unique potentials in accessing valley pseudo-spin degree of freedom (DOF)[2] and facile integration into quantum-photonic, electronic and sensing platforms via the layer-by-layer-assembly approach.[3] To date, QEs capable of operating in O-C telecommunication bands have not been demonstrated in TMDCs.[4-7] Here we report a deterministic creation of such telecom QEs emitting over the 1080 to 1550 nm wavelength range via coupling of 2D molybdenum ditelluride (MoTe$_2$) to strain inducing nano-pillar arrays.[8,9] Our Hanbury Brown and Twiss experiment conducted at 10 K reveals clear photon antibunching with 90% single photon purity. Ultra-long lifetimes, 4-6 orders of magnitude longer than that of the 2D exciton, are also observed. Polarization analysis further reveals that while some QEs display cross-linearly polarized doublets with ~1 meV splitting resulting from the strain induced anisotropic exchange interaction, valley degeneracy is preserved in other QEs. Valley Zeeman splitting as well as restoring of valley symmetry in cross-polarized doublets are observed under 8T magnetic field. In contrast to other telecom QEs,[10-12] our QEs which offer the potential to access valley DOF through single photons, could lead to unprecedented advantages in optical fiber-based quantum networks.**

Quantum emitters (QEs) that emit one photon at a time are key building blocks for numerous quantum technology protocols such as quantum communications, quantum information processing, and quantum key distribution.[13-15] In particular, single photon sources operated in the telecom bands (1.25–1.55 μm) are highly desired because they allow implementation of quantum technologies through existing fiber-based optical communication networks. An ideal QE should be on-demand, deterministic, and should deliver exactly one photon in a well-defined polarization and spatiotemporal mode.[16] Currently, telecom-compatible single photon sources have been demonstrated in various III-V semiconductor quantum dots[11,12] and recently in functionalized carbon nanotubes.[10,17] Although these single-photon sources have achieved telecom emission down to 1.55 μm, several challenges such as accurate site positioning and efficient polarization control still remain. Over the past decade, two-dimensional (2D) semiconductors have emerged as a novel optoelectronic platform for both fundamental research and advanced technological applications. Leveraged by the unique membrane-like planar geometry, 2D semiconductors are promising for QEs as they offer high photon extraction efficiency, easy coupling to external fields, and convenient integration with photonic circuits. In addition, as atomically-thin 2D flakes are stretchable and flexible, strain engineering can be readily applied to accurately position the emission sites.[3,8,9] Most importantly, atomically-thin 2D transition metal dichalcogenides (TMDCs) have a valley degree of freedom that can be manipulated and accessed through circularly polarized excitonic optical transitions and efficiently tuned via a magnetic field,[2,18] bringing new quantum functionalities to embedded QEs. Recently, 2D QEs have been demonstrated in $WSe_2$[4-7] and hexagonal boron nitride,[19,20] covering a broad emission spectrum range from ~500–800 nm. However, SPE emission in the most desirable spectral range – the telecom bands – has never been explored in 2D systems. While most 2D semiconductors have inherent electronic band structures that limit the operating wavelength to the visible range, alpha-phase 2H-MoTe$_2$ has a layer-dependent bandgap in the NIR regime, holding promise for telecom-compatible single photon emission.

We demonstrate the first observation of telecom single-photon emission in MoTe$_2$ mono- and few-layers. We transferred mechanically exfoliated MoTe$_2$ thin flakes onto nanopillar arrays to introduce point-like bi-axial strains, which locally trap excitons into the strain-defined potential landscape, leading to isolated emitters (see Methods and **Supplementary Section S1**). We observed quantum emitters in both monolayer MoTe$_2$ flakes and relatively thicker MoTe$_2$ samples up to >10 layers. **Figure 1a** is an optical image of a MoTe$_2$ monolayer on nanopillar arrays. The corresponding wide-field photoluminescence (PL) image, **Figure 1b**, shows the strained regions give significantly brighter PL emission than the flat areas. Unstrained monolayer MoTe$_2$ features two dominant emission peaks at cryogenic temperature: exciton ($X^0$) emission at around 1050 nm and trion ($X^{\pm}$) emission at around 1070 nm (**Figure 1c**).[21,22] With localized strain, a series of narrow PL peaks emerges from the lower energy side of the spectrum and covers a broad spectral range (**Supplementary Section S2**). The linewidths of such narrow PL emission lines range from a few meV to sub-meV at 10 K temperature, which are significantly narrower than the linewidths of the delocalized MoTe$_2$ exciton PL peaks, and are comparable to those of carbon nanotube (CNT)-based quantum emitters. The narrow near-band-edge emission lines are also frequently observed in 2–4 layered samples but not commonly seen in much thicker flakes. **Figure 1d** shows a typical narrow PL peak from a localized emitter that displays a linewidth (full width at half maximum, FWHM) of 920 μeV. The peak is accompanied by a weak shoulder peak at ~ 2.5 meV higher energy side. Both the width and shape of this PL peak remain essentially unchanged over nearly 3 order of magnitude change in pump power. The PL intensity also vary linearly with pump power over 2 order of magnitude change in pump power and only shows a weak saturation at powers >300 nW (**figure 1e**), indicating the robustness of the QE emission.

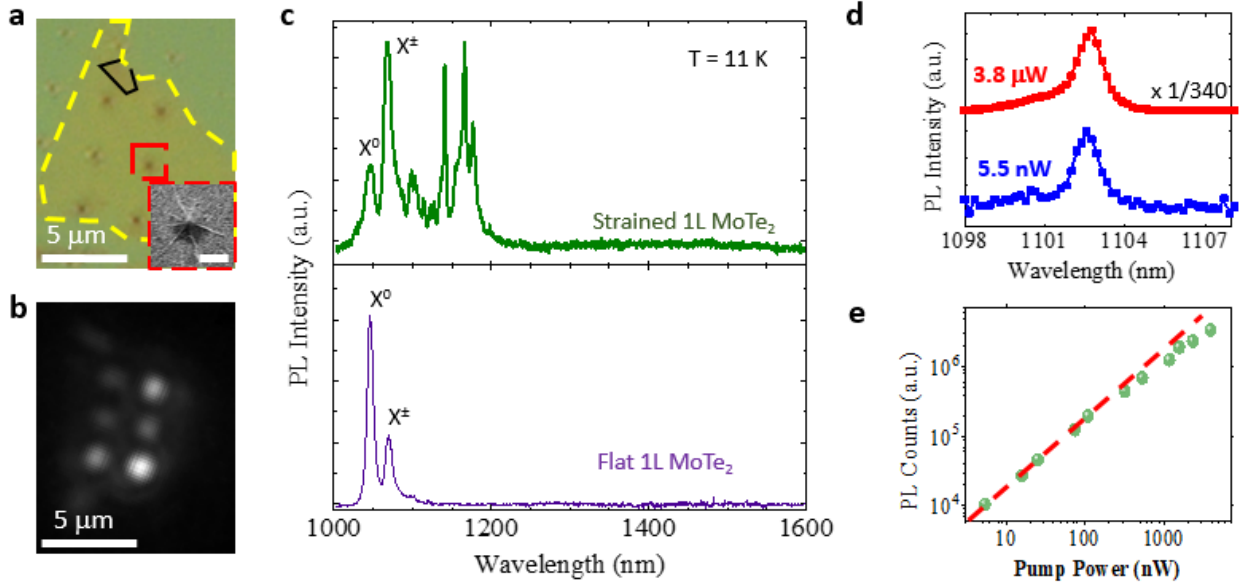

**Figure 1 | a**, Optical image of a monolayer MoTe$_2$ flake (with a tiny folded bilayer region outlined in black) on nanopillar arrays. The MoTe$_2$ flake is outlined by yellow dashed lines. Scale bar: 5 μm. Inset: SEM image of a nanopillar coated with monolayer MoTe$_2$, displaying a tent-shaped strain profile. Scale bar of the inset: 500 nm. **b**, Wide-field PL image of the same flake. The brighter areas, indicating stronger emission, are consistent with the pillar locations in **a**. Scale bar: 5 μm. **c**, PL spectrum of a MoTe$_2$ monolayer on flat PMMA (lower panel) and the PL spectrum of a MoTe$_2$ monolayer sitting on a nanopillar (upper panel). The MoTe$_2$ exciton and trion peaks are identified. **d**, PL spectra of a representative localized MoTe$_2$ emitter acquired at 3.3 μW and 5.5 nW pump power showing invariant shape and width (920 μeV at FWHM) of the spectral line. **e**, Pump-power-dependent emission intensity of the PL peak presented in **Figure 1d** shows near-perfect linear scaling with pump power. The red dash line is to guide the eye. All experimental data were taken at 11 K temperature.

To verify the single photon nature of the narrow emission peaks originated from localized strains, we performed Hanbury Brown-Twiss (HBT) experiments to measure photon antibunching. The second-order correlation at zero time delay, or g$^2$(0), was extracted to evaluate the probability of detecting two photons simultaneously. **Figure 2a** displays the PL spectrum of a strain-induced localized emitter, of which the PL dynamics and the auto-correlation measurement results are presented in **Figure 2b-d**. The time-resolved photoluminescence (TRPL) decay curve (**Figure 2b**) shows a near-perfect single-exponential decay with a lifetime τ = 22.2±0.1 ns that is four orders of magnitude longer than that in pristine $MoTe_2$ (~2 ps), which is attributed to the non-radiative decay dominated process.[23] This long lifetime provides a clear indication that localization of the exciton in a strain induced potential trap prevents the exciton from recombining via non-radiative defects that dominate the decay of 2D band-edge excitons.

Quantum-dot-like solid-state QEs typically have instability issues such as photon bleaching, blinking, and spectral diffusion, which hinders applications. For instance, the blinking and spectral diffusion behavior of PbS quantum dots (one material capable of quantum light emission at NIR range) is wellknown.[24] We monitored the time-dependent emission using both a single-photon detector (**Figure 2b, inset**) and an InGaAs spectrometer (**Supplementary Section S3**), finding no detectable photon bleaching, blinking, or spectral diffusion over the timescales presented. Some of our $MoTe_2$ quantum dots were measured continuously over 24 hours and no signs of degradation were detected. **Figure 2c and 2d** present the photon correlation under pulsed [$g^2(0) = 0.098±0.003$] and CW [$g^2(0) = 0.181±0.030$] excitation, respectively. Both values are well below the photon antibunching threshold of $g^2(0) = 0.5$, which unequivocally reveals the strain-induced $MoTe_2$ localized emitter is a QE.

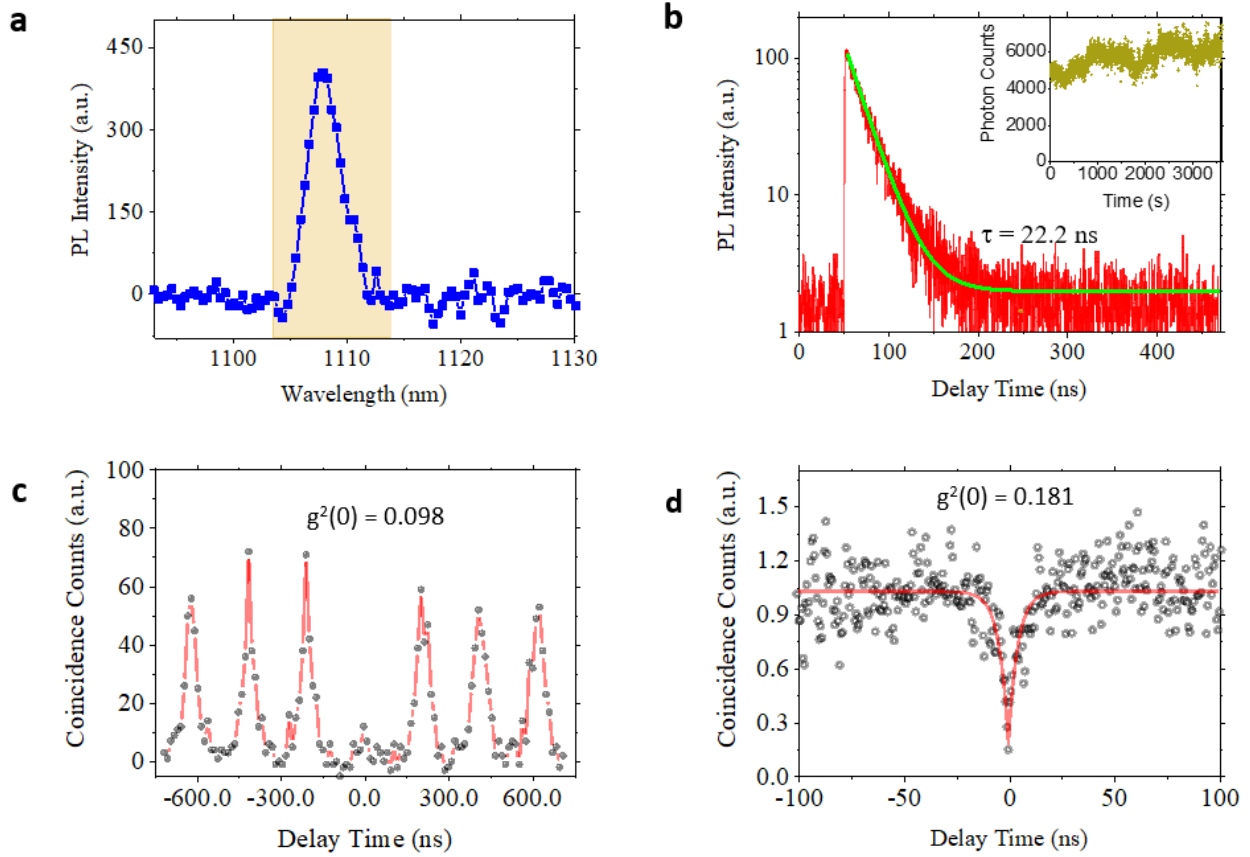

**Figure 2 | a**, PL spectrum of a localized MoTe$_2$ emitter. The data of **b, c, d** are taken from this dot with a band pass filter that allows the shadowed region to be detected. **b**, the PL decay curve (red) and a single exponential decay fit (green) with a 22.2±0.1 ns extracted lifetime. Inset plots the PL intensity as a function of experiment time showing stability of the single photon emission rate over 1 hour time. The modulation observed over the long time scale (~100 s) is mainly due to sample drift relative to the laser excitation spot. **c**, Second-order correlation measurement under a 850 nm pulse excitation with a 2.1 MHz repetition rate, from which a $g^2(0) = 0.098\pm0.003$ is extracted. **d**, Second-order correlation measurement under a 850 nm CW laser excitation. The red curve is a fit to the data using a bi-exponential decay function. The extracted $g^2(0) = 0.181\pm0.030$. All the experimental data were taken at 11 K temperature.

Obtaining telecom-compatible QEs that emit at around 1.3 μm (O-band) and 1.55 μm (C-band) are required for fiber-based quantum communications as the transmission loss in optical fibers are minimized in these bands. When increasing the MoTe$_2$ layer numbers from monolayer to bulk, the bandgap of MoTe$_2$ decreases from 1.18 eV (1050 nm) to 0.95 eV (1300 nm) monotonically.[22] The PL intensity also decreases orders of magnitude due to a direct-to-indirect bandgap transition commonly observed in 2D TMDCs. In our experiment, we observe bright telecom wavelength emissions created from strained few-layer MoTe$_2$ although we have occasionally found such emissions in mono- and bilayer samples (**Supplementary Section S4**). **Figure 3a** and **Figure S5a,b** present PL spectra of telecom-wavelength emitters. We typically observe such highly red-shifted bright emissions spanning 1.25–1.55 μm, covering the full telecom window. In contrast to QEs emitting at wavelengths below 1200 nm, these are characterized by relatively broad linewidths (FWHM 7-~30 meV). **Figure 3b** is the PL spectrum of a telecom emitter, of which PL dynamics and photon correlation results are presented in **Figure 3c-3f**, respectively. We observed an initial PL decay with a lifetime of 163 ± 3 ns, followed by an ultra-long decay lifetime of 1.13 ± 0.01 μs. This ultra-long lifetime is attributed to the 1540 nm emission, confirmed by comparing the integrated PL counts from the TRPL curve with the PL counts in spectrum (details given in **Supplementary Section S5**). The measured lifetime is 6 orders of magnitude longer than that of the MoTe$_2$ band-edge emission and is of 2–3 orders larger than that of the near-band-edge QEs mentioned before. Based on this long lifetime and the fact that these telecom-QEs are observed more on multilayer thick MoTe$_2$, we tentatively attribute the telecom-QEs to indirect excitonic transitions, which are activated by strain induced quantum confinement potentials. **Figure 3d** show that this telecom-QE is also free of blinking and photon bleaching over 5000 s experiment time. The pulsed excitation photon correlation measurement for the spectral window shown in **Figure 3b** yield a g$^2$(0) = 0.48±0.03 (**Figure 3e**). Since this measurement includes contribution from a high-energy shoulder that exhibits shorter PL decay, we employed a time gated g$^{(2)}$ experiment[25] (**Supplementary Section S6)**, in which only the photons arriving after the decay of the higher energy shoulder (i.e. after 200 ns delay) were analyzed for the $g^{(2)}$ trace. The time gated $g^{(2)}$ in **Figure 3f** shows $g^2(0) = 0.155\pm0.009$ clearly proving the antibunching nature of the 1540 nm telecom emitter. Our result is the first ever demonstration of a 2D material-based C-band (~1.55 μm) telecom QE.

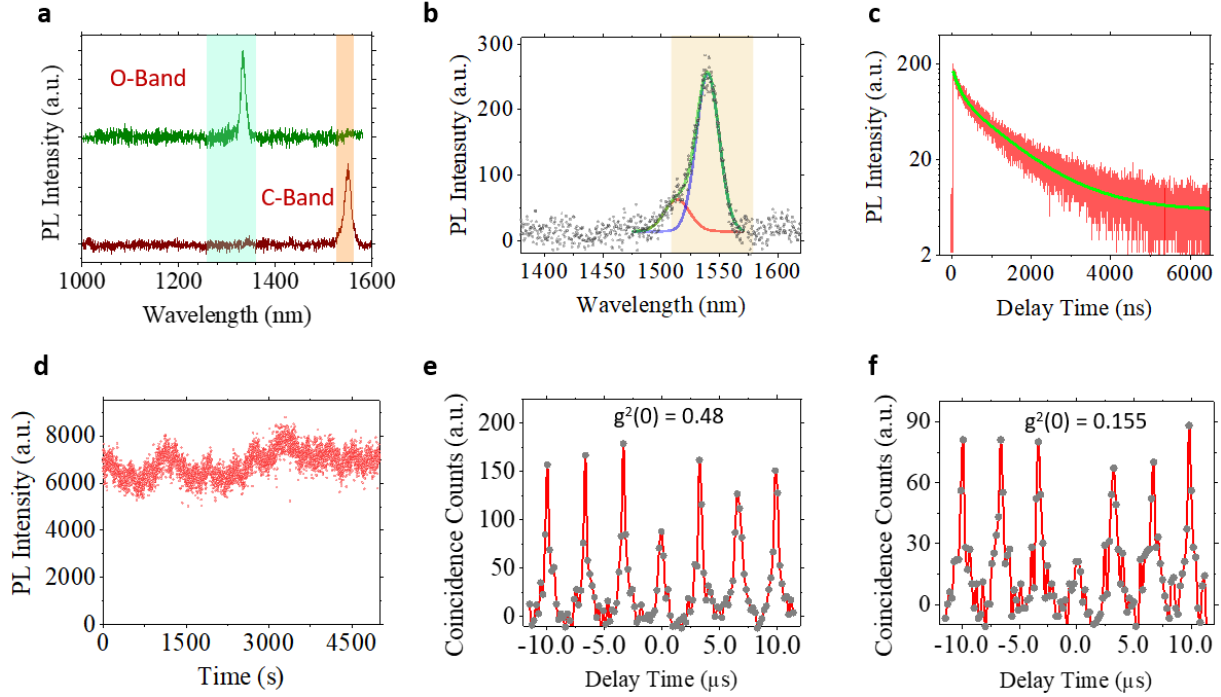

**Figure 3 | a**, "O-band" and "C-band" telecom PL emissions from two multilayer $MoTe_2$ localized emitters. **b**, $MoTe_2$ quantum dot with a 1540 nm telecom emission peak and a small shoulder peak at 1510 nm. The blue and red lines are Gaussian fits to the 1540 and 1510 nm peaks, respectively. The data of **c-f** are taken from this dot with a band pass filter that allows the shadowed region to be detected. **c**, PL decay curve (red) and a bi-exponential decay fit (green) reveal the lifetime of the 1540 nm peak is 1.13±0.01 μs. **d**, Time-dependent PL counts showing a stable emission over 5000 seconds. **e**, Second-order correlation measurement under a 850 nm pulsed excitation with a 330 kHz repetition rate, from which $g^2(0) = 0.48\pm0.03$ is extracted. **f**, Time-gated $g^{(2)}$ experiment shows $g^2(0) = 0.155\pm0.009$ under a 200 ns time gate. All experimental data were taken at 13 K temperature.

Semiconducting 2D TMDCs are well-known for their valley degree of freedom, which gives rise to interesting phenomena such as valley Zeeman splitting and valley polarization. To investigate the valley physics of the strain-induced QEs in MoTe$_2$, we conducted polarization-resolved magneto-PL spectroscopy with a magnetic field normal to the sample surface (Faraday geometry). **Figure 4a** presents a helicity-resolved PL study of a MoTe$_2$ QE. The spectra was taken with $\sigma^+$ excitation and analyzed for both $\sigma^+$ and $\sigma^-$ helicities. The valley Zeeman splitting was not observed in the absence of a magnetic field but rose with increasing field, indicating a lifting of valley degree degeneracy. Using the relation between the energy splitting $\Delta E$ and the magnetic field $B$, $\Delta E = -g\mu_B B$, where $\mu_B$ is the Bohr magneton, we extracted a Landé $g$-factor of -3.61±0.02 for the quantum dot (**Supplementary Section S7**), which is comparable with $g$-factors reported in other work.[26] At zero field, the degree of circular polarization, $P_C = (I_{\sigma+} - I_{\sigma-})/(I_{\sigma+} + I_{\sigma-})$, was 13% as a result of the valley selective-pumping effect. Significant valley polarization was observed with an applied magnetic field, reaching 52% at 8 T, which is more pronounced compared to the 30% polarization reported in magnetic-field induced valley polarization of intrinsic MoTe$_2$ excitons.[26]

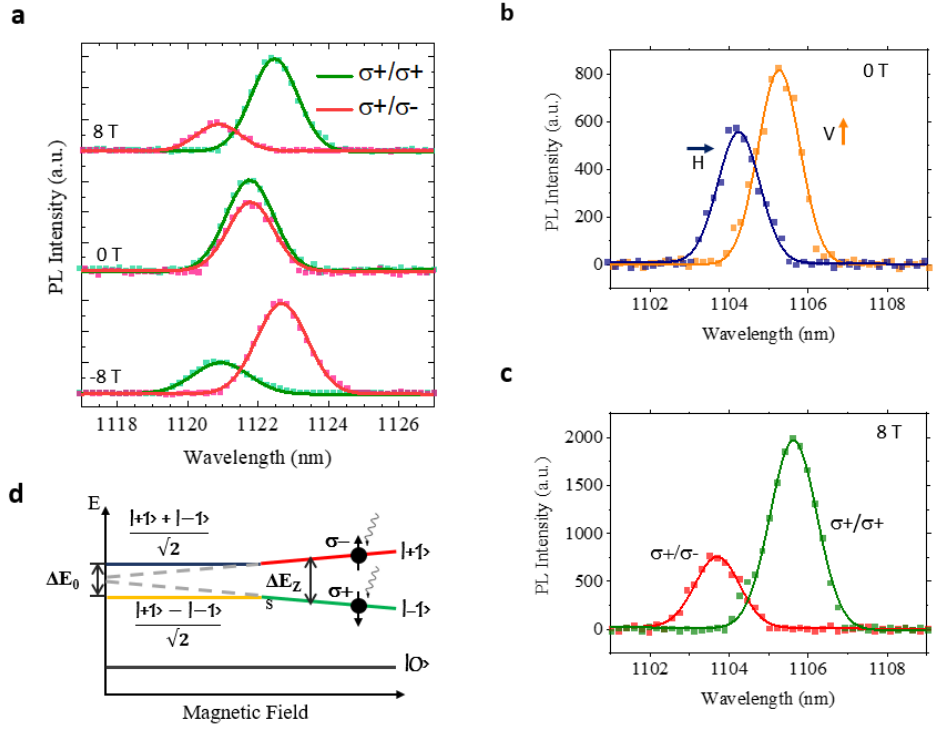

**Figure 4 | a**, Helicity-resolved magneto-PL data of a MoTe$_2$ localized emitter. The emitter was excited using a σ$^+$ polarized laser. **b**, The PL spectrum of a linearly cross-polarized doublet measured at zero field, showing a fine-structure energy splitting of ~1.1 meV. H and V denote horizontal and vertical polarization detection directions, respectively. **c**, Spectrum of the same doublet under an 8 T magnetic field, showing the doublet is converted to a circular cross-polarized pair. The excitation is a σ+ polarized laser. The lines in **a-c** are Gaussian peak fits to the PL data. **d**, Energy diagram of the doublet as a function of the external magnetic field. The linearly-polarized states are converted into σ$^+$ and σ$^−$ circularly polarized states once the Zeeman energy ($\Delta E_{ZS}$) exceeds the zero-field fine-structure splitting energy ($\Delta E_0$). The ground state is denoted as $|0\rangle$. All experimental data were taken at 11 K temperature.

In addition to the Zeeman splitting of valley degenerated trapped excitons, we also observed emission pairs in some of the QEs, which were found to be cross-linearly polarized doublets with sizeable zero-field energy splitting (1-3.7 meV) (**Figure 4b** and **Supplementary Section S8**). The two peaks from the doublet have a zero-field splitting of 1.09 meV and reach their maximal PL intensities in opposite linear polarization directions (horizontal/vertical), indicating the presence of fine structure. The observed cross-polarization and zero-field energy splitting have also been reported in III-V quantum dots[27-29] and recently in WSe$_2$ QEs.[4,6,30] Following prior studies, we attribute this fine structure splitting to hybridization of K and K' valley polarized excitons by an asymmetric potential landscape defined by the localized strain as illustrated in **Figure 4d.** In our system the zero-field splitting is larger than those (10s – 100s of µeV) observed in III-V quantum dots, indicating a strong electron-hole exchange interaction in such a strained 2D system. When a magnetic field is applied, the Zeeman energy splitting ($\Delta E_{ZS}$) and the anisotropic-potential-induced zero-field splitting ($\Delta E_0$) compete with each other. Once the field is strong enough to overcome the anisotropic Coulomb potential, that is, $\Delta E_{ZS} > \Delta E_0$, the linearly polarized states vanish and circularly polarized states are recovered. This restoration of valley symmetry is achieved in our QE under an 8 T magnetic field (**Figure 4c**). The helicities of each peak can be reversed by flipping the direction of the magnetic field, proving that the valley Zeeman effect is now dominating over anisotropic exchange.

In summary, we have deterministically created single-photon emitters in monolayer and multilayer MoTe$_2$ using nano-pillar-based strain engineering. The emission wavelength ranges from 1080–1550 nm, covering all telecom bands. The quantum nature of the localized emitters was verified by photon correlation measurements. We observed cross-polarized doublets with ~1.1 meV zero-field energy splitting in some of the localized emitters, suggesting strongly anisotropic confinement. The polarization of the QEs was found to be tunable by external magnetic field. Our findings extend the operating wavelength of 2D QEs into the NIR regime, bringing in new solutions for creating site-controlled stable telecom-compatible quantum emitters. We envisage various future directions inspired by our early-stage demonstration, including electrically driven telecom quantum emitters and cavity-enhanced tunable NIR QEs. We are also encouraged by the possibility of realizing room-temperature MoTe$_2$ telecom QEs as the energy redshift between the telecom emission and the MoTe$_2$ exciton emission is well above the thermal energy at room temperature. Finally, an in-depth understanding of the excitonic physics of MoTe$_2$ QEs may give rise to new perspectives on manipulating the spin-valley matrix and designing valleytronic devices.

**Methods**

**Sample Preparation.** MoTe$_2$ flakes were mechanically exfoliated from a flux-grown bulk crystal before they were transferred onto pre-patterned substrates. Thin layers (that is, flakes that look greenish and translucent under an optical microscope) were selected for further optical characterizations. 1-4 layer flakes can be easily distinguished by analyzing the band-edge emission wavelength and PL intensity.[22] Flakes thicker than four layers do not have a detectable band-edge emission using μW level pump power. To prepare the strained substrates, a 50 nm Au layer was deposited on top of a Si/SiO$_2$ substrate to block silicon emission, followed by the spin-coating of a 50 nm polyvinyl alcohol (PVA) dielectric layer to avoid quenching effects. Then, a ~120 nm polymethyl methacrylate (PMMA) layer was spin-coated on top of the PVA layer and patterned by electron beam lithography into PMMA nanopillar arrays with a 3 μm pitch width. A 90ºC vacuum annealing was applied to enhance the contact between the 2D flakes and the nanopillars. Each pillar had a ~100 nm pillar height and a ~150 nm diameter. Note: later we found the PVA spacer layer was not necessary as the ~100 nm PMMA pillars were sufficient to separate the MoTe$_2$ dots from the gold layer and maintain efficient PL emission.

**Optical Characterization.** A diagram of our optical measurement setup is presented in **Supplementary Section S9**. Micro-PL measurements of MoTe$_2$ QEs were performed on a home-built confocal microscope with excitation of either an 850 nm CW Ti:sapphire laser or an 850 nm supercontinuum pulsed laser. The excitation power was typically a few μW. Samples were mounted in a continuous flow cryostat and cooled to 10–13 °K using liquid helium. The emitted light was collected through a 50× infrared objective lens (Olympus, 0.65 NA) and spectrally filtered before entering a 2D InGaAs array detector (NIRvana 640LN, Princeton Instruments). We used 150 and 300 gr/mm gratings to resolve the spectra. For TRPL and HBT experiments, the emission signal was spectrally filtered before coupled into a 50:50 optical fiber beamsplitter, which equivalently split the signal into two beams and sent them into two channels of a superconducting nanowire single-photon detector (Quantum Opus). PL intensity time trace, PL decay curves and $g^2(\tau)$ traces were obtained from photon detection events recorded by a PicoQuant HydraHarp 400 time-correlated single photon-counting module. We applied a bi-exponential decay model to determine the CW $g^2(0)$ value and error level. For pulsed auto-correlation measurements, $g^2(0)$ was extracted by comparing the integrated photon coincidence counts at the zero-time delay peak with the averaged integrated photon coincidence counts at 30 adjacent peaks. The error level of pulsed $g^2(0)$ was defined by the standard deviation of the integrated photon coincidence counts in the adjacent peaks. For the pulsed $g^2(0)$ measurement of the 1540 nm peak, a time gate of 200 ns was applied to significantly reduce the contribution from the undesired emissions that could not be fully removed by optical filters.

For magneto spectroscopy and polarization-resolved PL measurements, the sample was placed inside the room temperature bore of an 8.5 T liquid-helium cooled superconducting magnet. For linear polarization analysis, the excitation beam was fully depolarized using a laser depolarizer to eliminate the polarization memory effect. A half-wave plate (HWP) was inserted into the collection channel to rotate the polarization direction. A Wollaston prism was placed between the HWP and the InGaAs detector to spatially split the emission into horizontal and vertical components, followed by a depolarizer to avoid effects arising from the linear polarization dependence of the gratings. For circular polarization analysis, a quarter-wave plate (QWP) was inserted into the shared path of the excitation and the emission beams. As a result, the QWP turns the linearly polarized laser into a circularly polarized excitation source and converts the circularly polarized PL signals into linearly polarized beams. A HWP and a Wollaston prism were installed in the collection beam to spatially split the $\sigma^+$ and $\sigma^-$ emissions into two different areas on the 2D InGaAs array.

## Acknowledgement

The authors would like to acknowledge the helpful discussion and technical support from Dr. Christopher Lane, Dr. Jianxin Zhu, Dr. Andrew Jones and Mr. John Kevin Scott Baldwin. This work was performed at the Center for Integrated Nanotechnologies, an Office of Science User Facility operated for the U.S. Department of Energy (DOE) Office of Science. Los Alamos National Laboratory (LANL), an affirmative action equal opportunity employer, is managed by Triad National Security, LLC for the U.S. Department of Energy's NNSA, under contract 89233218CNA000001. Deterministic quantum emitter creation capability was developed under the support of DOE BES, QIS Infrastructure Development Program BES LANL22. HZ, YZ and HH acknowledge partial support form Laboratory Directed Research and Development (LDRD) program 20200104DR. HH is also partially supported by Quantum Science Center. MTP is supported by LDRD 20190516ECR. HZ also acknowledge a partial support from LANL Director's Postdoctoral Fellow Award.

## Author contributions

HZ and HH conceived the experiment. HZ, under the supervision of HH, primarily developed deterministic QE creation approach, designed and conduct the experiment, analyzed the data and composed the paper. YZ assisted in the experiment. MTP assisted in the design of the nanopillar samples and paper preparation.

**Competing financial interests**

The authors declare no competing financial interests.